\documentclass[letterpaper, 10 pt, conference]{ieeeconf}  
\IEEEoverridecommandlockouts                           
\usepackage[usenames,dvipsnames]{xcolor}
\usepackage{algorithmic}
\usepackage{algorithm}
\usepackage{array}
\usepackage[caption=false,font=normalsize,labelfont=sf,textfont=sf]{subfig}
\usepackage{textcomp}
\usepackage{stfloats}
\usepackage{url}
\usepackage{verbatim}
\usepackage{graphicx}
\usepackage{cite}
\usepackage{amsmath,amssymb,amsfonts,mathtools,bm}
\usepackage{tikz}
\usepackage{siunitx}
\usepackage{arydshln}
\usepackage{balance}
\usepackage{lipsum}
\usepackage{booktabs}
\usepackage{bbold}
\usepackage{multirow}
\usepackage{marginnote}
\usepackage{color, soul}
\usepackage{multicol}
\usepackage[makeroom]{cancel}
\newtheorem{definition}{Definition}[section]

\usepackage{tabularx}
\usepackage{eurosym}
\usepackage{cases}

\usepackage{tikz}
\usetikzlibrary{spy}
\usepackage[utf8]{inputenc}
\usepackage{pgfplots}
\usepgfplotslibrary{groupplots,dateplot}
\usetikzlibrary{patterns,shapes.arrows}
\pgfplotsset{compat=newest}
\def\axisdefaultheight{110pt}

\DeclareMathOperator*{\argmin}{arg\,min}
\usepackage{eurosym}

\usepackage{hyperref}	% for making links clickable
\usepackage[labelfont=bf,labelsep=colon]{caption}
\begin{document}
\title{\LARGE \bf  
Model Predictive Control and Moving Horizon Estimation using Statistically Weighted Data-Based Ensemble Models
}

\author{Laura Boca de Giuli, Samuel Mallick, Alessio La Bella, Azita Dabiri, Bart De Schutter, Riccardo Scattolini
	\thanks{Laura Boca de Giuli, Alessio La Bella, and Riccardo Scattolini are with the Department of Electronics, Information, and Bioengineering, Politecnico di Milano, Italy (e-mails: \textsl{laura.bocadegiuli@polimi.it}, \textsl{alessio.labella@polimi.it},  and
	\textsl{riccardo.scattolini@polimi.it}). }
    \thanks{Samuel Mallick, Azita Dabiri and Bart De Schutter are with the Delft Center for Systems and Control, Delft University of Technology, The Netherlands (emails: \textsl{s.h.mallick@tudelft.nl}, \textsl{a.dabiri@tudelft.nl}, and \textsl{b.deschutter@tudelft.nl}.}
    }

\maketitle
\thispagestyle{empty}
\pagestyle{empty}

\begin{abstract}
  This paper presents a model predictive control (MPC) framework leveraging an ensemble of data-based models to optimally control complex systems under multiple operating conditions. A novel combination rule for ensemble models is proposed, based on the statistical Mahalanobis distance, enabling the ensemble weights to suitably vary across the prediction window based on the system input. In addition, a novel state observer for ensemble models is developed using moving horizon estimation (MHE). The effectiveness of the proposed methodology is demonstrated on a benchmark energy system operating under multiple conditions.
\end{abstract}


\section{INTRODUCTION}
\label{sec:intro}
Model predictive control (MPC) has become one of the most prevalent strategies for managing large-scale and complex systems, primarily thanks to its predictive capabilities and explicit constraints handling. Data-based models such as neural networks (NNs) have recently gained popularity in the MPC community due to the vast availability of data and the relative ease of development in comparison to physics-based models \cite{de2024modeling, mallick2025learning}. However, when diverse datasets spanning different operating ranges are used, employing a single black-box model to represent significantly varying operating conditions often leads to degraded performance and loss of prediction accuracy \cite{gawlikowski2023survey,de2024lifelong}. 

A promising solution to this issue is ensemble learning \cite{zhang2012ensemble}. This approach improves accuracy and generalisation by training specialised models, or experts, on specific datasets corresponding to different operating domains, and subsequently combining their outputs \cite{li2024theory}. Several techniques have been proposed in the literature to combine the predictions of ensemble models. One common approach is the arithmetic averaging of the models' outputs, as discussed in \cite{wu2019machine} for recurrent neural networks (RNNs). An alternative approach is weighted averaging, where higher weights are assigned to models whose predictions were more accurate in past data \cite{wang2022coscl}. However, as these methods select a static weighting, they may be suboptimal in predictive control frameworks such as MPC \cite{wu2019optimizing}, where the weighting should ideally vary across the prediction window. To this end, more advanced methods dynamically optimise the combination weights with gating networks \cite{jordan1994hierarchical}, e.g., using generic artificial neural networks \cite{yeganeh2022ensemble} or Bayesian neural networks \cite{li2020continual}. These approaches rely on previously encountered data to estimate combination weights as functions of system states, inputs, and outputs. This is, however, impractical within an MPC framework, where future state and output measurements are not available along the prediction window, thus preventing proper optimisation of the combination weights \cite{de2025ensemble}.

A further challenge in the use of ensemble models within MPC is the estimation of the state of each individual expert, given that only the overall system input and output are measurable. This problem is especially significant for data-based models in which the state retains no physical meaning. In \cite{peralez2022neural}, a Kazantzis-Kravaris-Luenberger observer is designed for each expert and subsequently combined, but the approach relies on assumptions valid only for specific model structures, excluding complex nonlinear models such as NNs. Other works investigate the use of ensembles of observers to improve state estimation for a single model \cite{petri2024hybrid}.
Moreover, state constraints must also be explicitly considered during the estimation phase, an essential aspect for most data-based models (e.g., amplitude-bounded hidden states in RNNs due to activation functions \cite{bonassi2022recurrent}).
However, to the best of our knowledge, no method has yet been proposed for constrained state estimation in ensemble models.

In light of the above discussion, this paper makes two main contributions. Firstly, we propose an MPC formulation that employs an ensemble model whose experts are combined using the statistical Mahalanobis distance \cite{montgomery2009statistical}, i.e., combination weights are assigned according to the statistical proximity between the current operating condition and the training input data of each expert. This allows the combination weights to be determined solely as a function of the input, whose future sequence is available in MPC frameworks. 
Secondly, we propose a novel observer design for ensemble models based on moving horizon estimation (MHE) \cite{allan2018moving}, which enables the correct estimation of the states of all experts in the ensemble while enforcing state constraints.

The proposed framework can be applied to the control of general dynamical systems, whose ensemble can be composed of any type of data-based model. In this work, we employ gated recurrent units (GRUs) within the family of RNNs, thanks to their strong approximation capabilities \cite{bonassi2022recurrent}. Simulations on a district heating system featuring multiple operating conditions \cite{krug2021nonlinear} demonstrate that the proposed Mahalanobis-based ensemble MPC significantly outperforms standard averaging and optimisation-based methods of weight selection in terms of both economic performance and constraint satisfaction. Moreover, the MHE-based observer significantly reduces the output prediction error with respect to the ensemble model's open-loop predictions.

\section{Notation and preliminaries}
\label{sec:preliminaries}
Let $\mathbb{N}$ denote the set of natural numbers, $\mathbb{R}$ the set of real numbers, and $\mathbb{R}_{\geq 0}$ the set of non-negative real numbers. Given a vector $z \in \mathbb{R}^n$, its transpose is denoted as $z^{\top}$ and its $i$-th element as $z_i$. A vector of $n$ ones is denoted by $\bm{1}_n$. For a vector $z \in \mathbb{R}^n$ observed over $m$ time steps, i.e., $z(1), \hdots, z(m)$, where the set of time indices is $\mathcal{I}=\{1,\hdots,m\}$ with cardinality $|\mathcal{I}|$, the matrix containing these $m$ observations is denoted in bold as $\bm{z}=\big[z(1) \hdots z(m)\big] \in \mathbb{R}^{n \times m}$ and can be compactly expressed as $\bm{z}=\big\{z(k)\big\}_{ k \in \mathcal{I}}$. Matrices containing different vectors observed over $m$ time steps included in $\mathcal{I}$, e.g., $\bm{z}_1, \bm{z}_2$, are generically indicated using calligraphic letters, i.e., $ \mathcal{D} = \big[ \bm{z}_1^{\top} \, \bm{z}_2^{\top} \big]^{\top} $.

In this paper, we make use of the \textit{Mahalanobis distance} \cite{montgomery2009statistical}, a statistical process control metric used to quantify how far two datasets are from each other in terms of mean and covariance. 

\begin{definition} \label{def:T2} (\textbf{Mahalanobis distance}).
	Consider a benchmark dataset $\bm{z}=\big\{z(k)\big\}_{ k \in \mathcal{I}}$ containing $|\mathcal{I}|$ observations of a variable $z \in \mathbb{R}^{n_z}$, and a monitoring dataset \mbox{$\bm{\widetilde{z}}=\big\{\widetilde{z}(k)\big\}_{ k \in \widetilde{\mathcal{I}}}$} with $|\widetilde{\mathcal{I}}|$ observations of $\widetilde{z} \in \mathbb{R}^{n_z}$, where $\mathcal{I}$ and $\widetilde{\mathcal{I}}$ contain the time indices of the observations in $\bm{z}$ and $\bm{\widetilde{z}}$, respectively. The statistical Mahalanobis distances $T^2$ of $\bm{\widetilde{z}}$ with respect to $\bm{z}$ are defined as 
	\begin{equation}
		\label{eq:T2}
		\begin{aligned}
			T^{2}(\bm{\widetilde{z}},\bm{z}) = \Big\{\big(\widetilde{z}(k)-\mu_{\bm{z}}\big)^{\top} (\Sigma_{\bm{z}})^{-1} \big(\widetilde{z}(k)-\mu_{\bm{z}}\big)\Big\}_{ k \in \widetilde{\mathcal{I}}}^{\top}\,,
		\end{aligned}
	\end{equation}
	where $\mu_{\bm{z}} = \frac{1}{|\mathcal{I}|}\sum\limits_{ k \in \mathcal{I}} \; z(k)$ and $\Sigma_{\bm{z}} = \frac{1}{|\mathcal{I}|-1}\sum\limits_{ k \in \mathcal{I}} \; \big(z(k) - \mu_{\bm{z}}\big)\big(z(k) - \mu_{\bm{z}}\big)^{\top}$. In detail, $T^{2}(\bm{\widetilde{z}},\bm{z}) \in \mathbb{R}^{|\widetilde{\mathcal{I}}|}_{\geq 0}$ contains a sequence of Mahalanobis distances, computed for each observation in $\widetilde{\bm{z}}$ with respect to the benchmark dataset $\bm{z}$. 
\end{definition}

\section{Problem statement}
\label{sec:problemstatement}
Consider a process modelled as a discrete-time system $\mathcal{S}$, with sampling time $\tau$,
\begin{equation}
	\label{eq:system}
	\mathcal{S}: \; \left\{
	\begin{aligned}
		x_{\text{p}}(k+1) &=\; f_{\text{p}}\big(x_{\text{p}}(k), u(k)\big) \\
		y_{\text{p}}(k) \;\;&=\; g_{\text{p}}\big(x_{\text{p}}(k), u(k)\big) 
	\end{aligned},
	\right.
\end{equation}
with $f_{\text{p}}$ and $g_{\text{p}}$ generic functions, $x_{\text{p}} \in \mathbb{R}^{n_{x_{\text{p}}}}$, $u \in \mathbb{R}^{n_u}$, and $y_{\text{p}} \in \mathbb{R}^{n_y}$ the state, input, and output vectors of the process, respectively, and $k$ a discrete-time counter for time steps of length $\tau$. Specifically, the input $u=\big[\tilde{u}^{\top} d^{\top}\big]^{\top}$ includes all exogenous signals acting on $\mathcal{S}$, i.e., manipulated control variables $\tilde{u} \in \mathbb{R}^{n_{\tilde{u}}}$ and external disturbances $d \in \mathbb{R}^{n_{d}}$, which are assumed to be measurable and predictable. The system is assumed to operate under multiple conditions, each characterised by distinct datasets representing system behaviour in those conditions. A typical example can be observed in energy systems, where load demand shifts between seasonal operating ranges, such as transitioning from summer to winter conditions.

To manage this system, we employ the well-established model predictive control (MPC) approach. As MPC is designed under the certainty equivalence principle, the performance strongly depends on the accuracy of the underlying model. For the systems under consideration in this paper, where vast amounts of data across multiple and heterogeneous operating regions are available, it is often more effective to use an ensemble of data-based models for prediction, rather than relying solely on a single physics-based or black-box model \cite{zhang2012ensemble}. Let us denote these models, indexed with $i \in \mathcal{N} = \{1,\dots,n\}$, as $\mathcal{M}^{\scriptscriptstyle[1]},\hdots,\mathcal{M}^{\scriptscriptstyle[n]}$, each trained independently on a specific dataset $\mathcal{D}^{\scriptscriptstyle[1]},\hdots,\mathcal{D}^{\scriptscriptstyle[n]}$ corresponding to a different operational domain. The $i$-th dataset is defined as $\mathcal{D}^{\scriptscriptstyle[i]} = \big[ \bm{u}^{\scriptscriptstyle[i] \top} \: \bm{y}_{\text{p}}^{\scriptscriptstyle[i] \top} \big]^{\top}$, where \mbox{$\bm{u}^{\scriptscriptstyle[i]}=\big\{u(k)\big\}_{ k \in \mathcal{I}^{\scriptscriptstyle[i]}}$} and  \mbox{$\bm{y}_{\text{p}}^{\scriptscriptstyle[i]}=\big\{y_{\text{p}}(k)\big\}_{ k \in \mathcal{I}^{\scriptscriptstyle[i]}}$} are the input-output data collected at time indices within the set $\mathcal{I}^{\scriptscriptstyle[i]}$. The outputs of the individual models are then combined to yield the overall output of the ensemble prediction model $\mathcal{M}$, expressed as
\begin{subequations}
	\label{eq:ensemble}
	\begin{numcases}{\mathcal{M}:}
		\mathcal{M}^{\scriptscriptstyle[i]}: \;
		\begin{cases}
			\label{eq:modeli}
			x^{\scriptscriptstyle[i]}(k+1) = f^{\scriptscriptstyle[i]}\big(x^{\scriptscriptstyle[i]}(k), u(k)\big) \label{eq:state_eq} \\
			y^{\scriptscriptstyle[i]}(k) = g^{\scriptscriptstyle[i]}\big(x^{\scriptscriptstyle[i]}(k), u(k)\big)
		\end{cases} \\
		\begin{aligned}
			\label{eq:ycomb}
			\qquad\qquad y(k) = \sum\limits_{i=1}^n \lambda^{\scriptscriptstyle[i]}(k) y^{\scriptscriptstyle[i]}(k)
		\end{aligned}
	\end{numcases}
\end{subequations}
where $f^{\scriptscriptstyle[i]}$ and $g^{\scriptscriptstyle[i]}$ are generic parametric functions characterising model $\mathcal{M}^{\scriptscriptstyle[i]}$, whose parameters are identified from the dataset $\mathcal{D}^{\scriptscriptstyle[i]}$.
Furthermore, \mbox{$x^{\scriptscriptstyle[i]} \in \mathbb{R}^{n_{x}^{\scriptscriptstyle[i]}}$}, $y^{\scriptscriptstyle[i]} \in \mathbb{R}^{n_{y}}$ are the state and output variables of the $i$-th model, respectively, and $y \in \mathbb{R}^{n_{y}}$ is the output of the ensemble obtained through a convex combination of $y^{\scriptscriptstyle[i]}$, defined by $\lambda^{\scriptscriptstyle[i]} \in \mathbb{R}_{\geq 0}$, for $i = 1,\dots,n$. Two main challenges arise when employing a data-based ensemble prediction model within an MPC framework:
\begin{enumerate}
    \item determining an appropriate rule to combine the outputs of the individual models through the weights $\lambda^{\scriptscriptstyle[i]}$, which, for optimality, should vary along the prediction window as a function of available variables;
    \item estimating the states of the data-based ensemble model while enforcing state constraints.
\end{enumerate}

\section{MPC with Mahalanobis distance-based ensemble model}
\label{sec:mpc}
The first contribution of this work is an MPC regulator that uses a data-based ensemble model for prediction, where the outputs of each ensemble component are combined according to the statistical proximity between the optimised inputs and those used to train the specific model. This strategy is inspired by \cite{de2025ensemble,de2025learning}, where, however, control design and state estimation for ensembles are not addressed. Considering a sampling time $\tau$ and a prediction horizon $H$, at each time instant $t=k \tau$, with $k \in \mathbb{N}$, the following MPC problem is solved according to the receding horizon strategy:
{
\setlength{\abovedisplayskip}{0pt}
\setlength{\abovedisplayshortskip}{0pt}
\begin{subequations}
	\label{eq:mpc}
	\begin{align}
		& P_\text{\tiny MPC}\big(\{\hat{x}^{\scriptscriptstyle[i]}(k)\}_{i \in \mathcal{N}}, \bm{d}\big) = \min_{\tilde{u}(h)}   \sum_{h = k}^{k+H-1}  \mathcal{L}(h)  \label{subeq:cost} \\
        & \textrm{s.t.} \:\:\,\, \forall h \in \{k,\hdots,k+H-1\} \nonumber \\ 
        & \qquad x^{\scriptscriptstyle[i]}(h+1) = f^{\scriptscriptstyle[i]}\big(x^{\scriptscriptstyle[i]}(h), u(h)\big), \: i \in \mathcal{N}, \label{subeq:xi} \\
        & \qquad y^{\scriptscriptstyle[i]}(h) = g^{\scriptscriptstyle[i]}\big(x^{\scriptscriptstyle[i]}(h), u(h)\big), \: i \in \mathcal{N}, \label{subeq:yi} \\
        & \qquad y(h) = \sum\limits_{i=1}^n \lambda^{\scriptscriptstyle[i]}(h) y^{\scriptscriptstyle[i]}(h), \label{subeq:y} \\
        & \qquad x^{\scriptscriptstyle[i]}(k) = \hat{x}^{\scriptscriptstyle[i]}(k), \: i \in \mathcal{N},  \label{subeq:x0} \\
        & \qquad x^{\scriptscriptstyle[i]}(h+1) \in \mathcal{X}, \: i \in \mathcal{N}, \label{subeq:xconstr} \\
        & \qquad \tilde{u}(h) \in \mathcal{U}, \label{subeq:uconstr} \\
        & \qquad y(h) \in \mathcal{Y}(h),\label{subeq:yconstr}
	\end{align}
\end{subequations}
}
\!\!where $\bm{d} = \big[d^\top(k) \dots d^\top(k+H-1)\big]^\top$ are known estimates of the external disturbances that enter the model input, i.e., $u(h) = \big[\tilde{u}(h)^\top d^\top(h)\big]^\top$.
In detail, the cost function \eqref{subeq:cost} represents a generic objective such as economic management or setpoint tracking. Constraint \eqref{subeq:x0} imposes the initial state for each model in the ensemble (see Section \ref{sec:mhe}). Constraints \eqref{subeq:xconstr}-\eqref{subeq:uconstr} enforce the states and control variables to lie within the admissible sets $\mathcal{X}$ and $\mathcal{U}$, respectively.
Similarly, constraint \eqref{subeq:yconstr} enforces the output to lie within $\mathcal{Y}(h)$, where the time dependency accounts for time-varying limits.
The dynamical model of the system \eqref{eq:ensemble} is embedded in the MPC formulation through constraints \eqref{subeq:xi}-\eqref{subeq:y}. 

As discussed in Section \ref{sec:intro}, existing MPC formulations embedding ensemble models leverage suboptimal combination rules for model outputs. To overcome this limitation, we propose a strategy in which the ensemble weights are designed as a function of the input $u$, and are optimised implicitly along the prediction window, i.e., 
\begin{equation}
	\label{eq:lambdai}
	\begin{aligned}
		\lambda^{\scriptscriptstyle[i]}_{\text{MD}}\big(u(h)\big) = \frac{\frac{1}{T^2\big(u(h), \bm{u}^{\scriptscriptstyle[i]}\big)}}{\sum\limits_{i=1}^n \frac{1}{T^2\big(u(h), \bm{u}^{\scriptscriptstyle[i]}\big)}}
	\end{aligned},
\end{equation}
where \mbox{$T^2\big(u(h), \bm{u}^{\scriptscriptstyle[i]}\big)$} is the Mahalanobis distance\footnote{In practice, an arbitrarily small constant is added to the denominators to avoid division by zero.}, computed according to Definition \ref{def:T2}, between the current input $u(h)$ and the training inputs of model $\mathcal{M}^{\scriptscriptstyle[i]}$.
With this approach, the smaller the statistical distance \mbox{$T^2\big(u(h), \bm{u}^{\scriptscriptstyle[i]}\big)$} between the input $u(h)$ and the training inputs $\bm{u}^{\scriptscriptstyle[i]}$ of $\mathcal{D}^{\scriptscriptstyle[i]}$, the larger the weight assigned to the corresponding model output $y^{\scriptscriptstyle[i]}$. This combination rule offers several advantages over traditional strategies within an MPC framework, as will be shown in Section \ref{sec:results}. The weights are able to vary along the prediction window, thus allowing the MPC to account for operating conditions varying within the prediction window, improving closed-loop performance. Furthermore, since the weights $\lambda^{\scriptscriptstyle[i]}_{\text{MD}}$ depend solely on the inputs, rather than on unavailable future measured states and outputs, the proposed combination rule is reliably computable over the whole prediction horizon $H$.
Finally, an input-based selection is further justified by the fact that, for different operating conditions, inputs vary significantly, while states and outputs may remain nearly unchanged, as they are regulated in the closed loop.
Note that, although \eqref{eq:lambdai} adds further complexity to the MPC problem, in Section \ref{sec:results} the computational burden of \eqref{eq:mpc} is shown to remain reasonable.

\section{MHE-based observer for ensemble models}
\label{sec:mhe}
The second contribution of this work is an observer scheme for estimating the state of each model $\mathcal{M}^{\scriptscriptstyle[i]}$ in the ensemble \eqref{eq:ensemble}.
In data-based models, the state of the prediction model often loses any physical significance, e.g., the internal state of an RNN, rendering state estimation essential.
To address this, we here propose relying on moving horizon estimation (MHE) \cite{allan2018moving}.
As a moving-horizon and optimisation-based observer method, MHE is a natural counterpart for MPC controllers, allowing the use of the same MPC prediction model for state estimation.
In particular, in our setting, MHE allows us to embed the data-based model of each ensemble component in the observer scheme.
Furthermore, state constraints can be handled explicitly.

For a prediction horizon $\hat{H}$, at each time instant $t=k \tau$, with $k \in \mathbb{N}$, the following MHE problem, a function of the previous $\hat{H}$ applied inputs and measured outputs, $\bm{u} \in \mathbb{R}^{n_u \times \hat{H}}$ and $\bm{y}_{\text{p}} \in \mathbb{R}^{n_y \times \hat{H}}$, is solved for each model in the ensemble, according to the receding horizon strategy:
{
\setlength{\abovedisplayskip}{5pt}
\setlength{\abovedisplayshortskip}{1pt}
\begin{subequations}
	\label{eq:mhe}
	\begin{align}
		P^{\scriptscriptstyle[i]}_\text{\tiny MHE}\big(&\bar{x}^{\scriptscriptstyle[i]}, \bm{u}, \bm{y}_{\text{p}} \big) =  \nonumber \min_{\bm{\hat{x}}^{\scriptscriptstyle [i]}, \bm{\omega}^{\scriptscriptstyle [i]}, \bm{\nu}^{\scriptscriptstyle [i]}} \big\|\hat{x}^{\scriptscriptstyle [i]}(k - \hat{H}) - \bar{x}^{\scriptscriptstyle [i]}\big\|_{Q_{x}} \nonumber \\
        & \qquad\qquad+ \sum_{h = k - \hat{H}}^{k-1}  \big\|\nu^{[i]}(h)\big\|_{Q_\nu} + \big\|\omega^{[i]}(h)\big\|_{Q_\omega} \label{subeq:costmhe} \\
        & \textrm{s.t.} \:\:\,\, \forall h \in \{k- \hat{H},\hdots,k-1\}\nonumber \\ 
        & \qquad \hat{x}^{\scriptscriptstyle[i]}(h+1) = f^{\scriptscriptstyle[i]}\big(\hat{x}^{\scriptscriptstyle[i]}(h), u(h)\big) + \omega^{[i]}(h), \label{subeq:ximhe} \\
        & \qquad y_{\text{p}}(h) = g^{\scriptscriptstyle[i]} \big(\hat{x}^{\scriptscriptstyle[i]}(h), u(h)\big) + \nu^{[i]}(h), \label{subeq:yimhe} \\
        & \qquad \hat{x}^{\scriptscriptstyle [i]}(h) \in \mathcal{X}, \label{subeq:mheX}
	\end{align}
\end{subequations}
}
\!\!where $\omega^{[i]}$ and $\nu^{[i]}$ are the process and measurement error, with the matrices $Q_\omega$ and $Q_\nu$ weighting their penalties, respectively.
Furthermore, $\bar{x}^{\scriptscriptstyle [i]}$ is the state estimate of \eqref{eq:mhe} for time step $k - \hat{H}$, computed at time step $k-1$.
The first cost term, the arrival cost, penalises deviation from this previous estimate, weighted by $Q_x$, in order to capture historical data outside the prediction window.
Finally, $\mathcal{X}$ in \eqref{subeq:mheX} is a constraint set imposed on the estimated state for, e.g., enforcing amplitude bounds on the hidden states in RNNs that arise from specific activation functions.

With \eqref{eq:mhe}, we propose to estimate the state of each model within the ensemble by using the global output of the real plant $y_{\text{p}}$ to form an output error.
The optimisation problem \eqref{eq:mhe} finds a state trajectory for each model in the ensemble that best matches the historical data of the true system.
The final optimised states $\hat{x}^{\scriptscriptstyle[i]}(k)$ are then used as the initial states in \eqref{subeq:x0}.
The combination of \eqref{eq:mhe} and \eqref{eq:mpc} can be interpreted as first pushing each model in the ensemble to best match the real system in the past, and then, as not all models will accurately capture the true dynamics for a given operating condition, weighting the contribution of each model in the future via \eqref{eq:lambdai}.

\section{Numerical Results}
\label{sec:results}
\begin{figure}[t!]
	\centering
	\includegraphics[width=0.4 \textwidth]{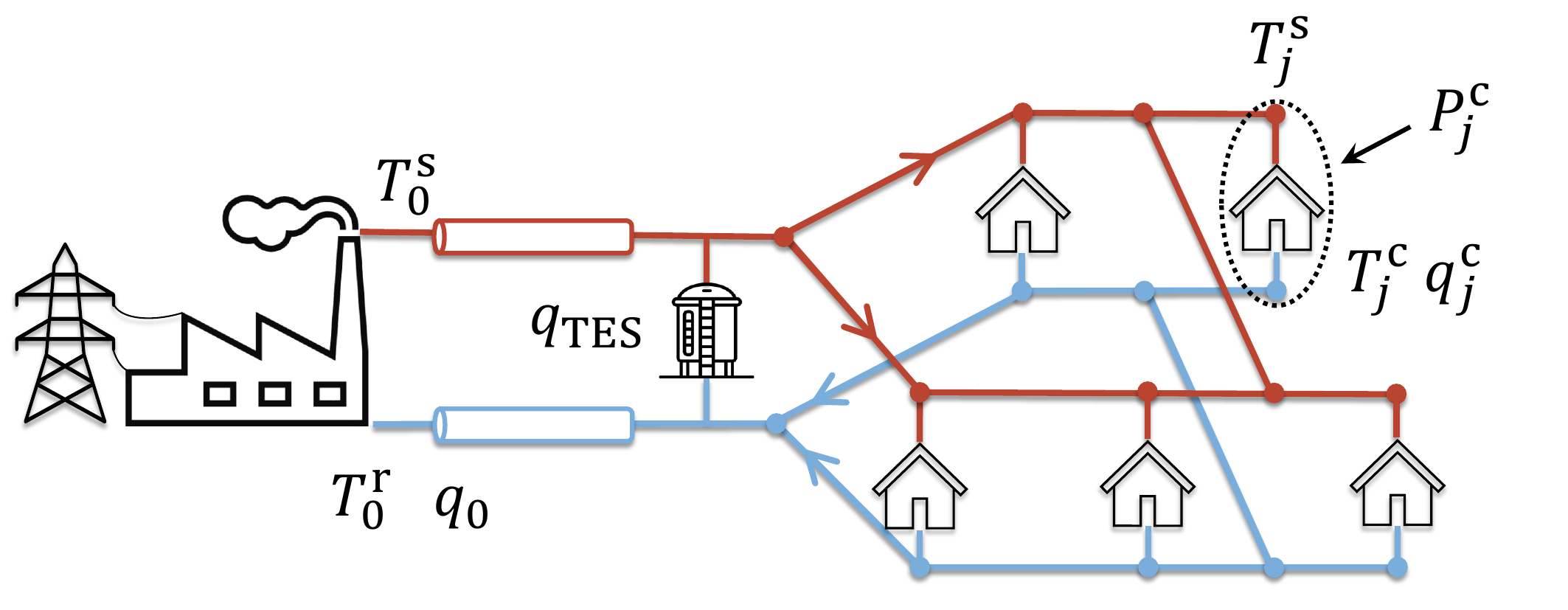}
	\caption{Schematic representation of the AROMA DHS and its variables.}
	\label{fig:aroma}
\end{figure}

The proposed framework is tested on a district heating system (DHS), a large-scale energy system consisting of a heating station with multiple thermal generators and an insulated water pipeline network that transfers heat to thermal loads. The DHS can also contain thermal energy storages, i.e., insulated water tanks whose water flow can be controlled to store or release energy \cite{nigro2024control}. 
The case study analysed in this work considers the AROMA DHS, presented in \cite{krug2021nonlinear} and depicted in Figure \ref{fig:aroma}. A high-fidelity dynamic simulator based on a dedicated Modelica library \cite{alvarado2024development} serves as a digital twin to generate input-output data for training the data-based models and to act as the real physical system for control. 

The following system variables are considered as inputs and outputs for the data-based model, as shown in Figure \ref{fig:aroma} and described in \cite{de2024physics}. The control variables are the supply temperature at the heating station, denoted by \mbox{$T_0^\text{s}$}, and the water flow from the thermal storage, denoted by $q_\text{\tiny TES}$.
The external disturbances are the five thermal load demands, i.e., $P_j^\text{c}$ for $j=1,\hdots,5$, with the overall input vector thus defined as $u = [T_0^\text{s} \; q_\text{\tiny TES} \; P_1^\text{c} \, \hdots \, P_5^\text{c} ]^{\top} $. 
The output variables include the return temperature $T_0^\text{r}$ and the water flow $q_0$ at the heating station, as well as the supply temperature $T_j^\text{s}$, output temperature $T_j^\text{c}$, and water flow $q_j^\text{c}$ for each $j$-th thermal load. The corresponding output vector is thus given by \mbox{$y_\text{p} = [ T_0^\text{r} \; q_0 \; T_1^\text{s} \, \hdots \, T_5^\text{s} \; T_1^\text{c} \, \hdots \, T_5^\text{c} \; q_1^\text{c} \, \hdots \, q_5^\text{c}]^{\top}$}. All temperatures are expressed in [$^\circ$C], all water flow rates in [kg/s], and all powers in [W]. 
Following the methodology presented in this work, an ensemble of data-based models is used to identify the AROMA DHS network, with the thermal energy storage modelled analytically as in \cite{nigro2024control}. 

For this case study, operational data from the DHS are assumed to be attainable for two working conditions: one characterised by lower power demand and the other by higher demand. These two operation modes can represent the summer and winter operating conditions of a DHS \cite{la2023optimal}.
Consequently, two data-based models are identified to form the ensemble. For the first working condition, a one-week input-output dataset $\mathcal{D}^{\scriptscriptstyle[1]}$ is collected with a sampling time of $\tau = 5$ minutes to identify model $\mathcal{M}^{\scriptscriptstyle[1]}$, where the controllable supply temperature $T_0^\text{s}$ is varied across its entire operating range using multilevel pseudorandom binary sequences (MPRBS) and the disturbances $P_j^\text{c}$, for $j=1,\hdots,5$, follow typical profiles observed in DHSs \cite{la2023optimal}, assumed to be measured during a specific season with thermal demand between 100 and 300 kW \cite{de2025learning}. A second one-week input-output dataset $\mathcal{D}^{\scriptscriptstyle[2]}$ is then collected to identify model $\mathcal{M}^{\scriptscriptstyle[2]}$, again varying $T_0^\text{s}$ over its full range using MPRBS, while considering the disturbance profiles representative of a different season with thermal demand between 150 and 350 kW \cite{de2025learning}. Although various model architectures could be employed, in this work we adopt the data-based model proposed in \cite{de2024physics}, which employs gated recurrent unit (GRU) networks within the recurrent neural network (RNN) family, due to their strong approximation capabilities and simple architecture \cite{bonassi2022recurrent}.

The MPC controller \eqref{eq:mpc} is implemented with the described ensemble as prediction model, horizon $H = 72$, and cost $\mathcal{L}(h) = \tau \cdot c(h) \cdot P_0(h)$, with $c(h)$ the electricity energy cost and $P_0(h) = c_\text{p} \cdot q_0(h) \cdot \big(T_0^\text{s}(h) - T_0^\text{r}(h)\big)$ the power supplied by the heating station, where $c_\text{p}$ is the specific water heat coefficient.
Furthermore, the constraint sets are defined as
\begin{equation}
    \begin{aligned}
        \mathcal{X} &= \{x \in \mathbb{R}^{n_x} | -\bm{1}_{n_x} \leq x \leq \bm{1}_{n_x} \}, \\
        \mathcal{U} &= \bigg\{\tilde{u} \in \mathbb{R}^{n_{\tilde{u}}} |65 \leq T_0^{\text{s}} \leq 85, -15 \leq q_\text{\tiny TES} \leq 15 \bigg\}, \\
        \mathcal{Y}(h) &= \big\{y \in \mathbb{R}^{n_y} | 2 \leq q_0^\text{r} \leq 25, \: \underline{T}_0^\text{r}(h) \leq T_0^\text{r} \leq 75, \\
        &\quad\quad \underline{T}_j^\text{s}(h) \leq T_j^\text{s} \leq 85, j=1,\hdots,5\big\},
    \end{aligned}
\end{equation}
where $\underline{T}_0^\text{r}$ and $\underline{T}_j^\text{s}$ are time-varying lower temperature bounds, e.g., being higher during the day and lower at night, consistently with the daily trend of the thermal demand \cite{la2023optimal}.
Note that the output constraints are `softened' with slack variables that are penalised in the cost to ensure feasibility in the presence of model errors. Finally, for the MHE-based observer \eqref{eq:mhe}, the selected horizon is $\hat{H} = 50$ and the weight matrices are identity matrices of appropriate dimension.

Four strategies are compared for selecting the weights $\lambda^{\scriptscriptstyle[i]}(h)$ for $h=k,\dots,k+H-1$ in the MPC controller:
\begin{itemize}
    \item \textbf{AV}: the average of the ensemble outputs is applied, i.e., $\lambda^{\scriptscriptstyle[i]}(h) = 1/n, \,\, \text{for} \,\, i = 1,\dots,n$.
    \item \textbf{LS}: the weights are optimised in a least-squares sense with respect to the last $100$ time steps of observations:
    \begin{equation}
    \begin{aligned}
    \{\lambda^{\scriptscriptstyle[i]}(h)\}_{i \in \mathcal{N}} = &\argmin_{\{\lambda^{\scriptscriptstyle[i]}\}_{i \in \mathcal{N}}} \sum_{h^\prime = k-100}^{k-1} \Bigg\|y_{\text{p}}(h^\prime) \\
    & -\sum_{i \in \mathcal{N}} \lambda^{\scriptscriptstyle[i]} g^{\scriptscriptstyle[i]} \big(x^{\scriptscriptstyle[i]}(h^\prime), u(h^\prime)\big) \Bigg\|^2. \nonumber
    \end{aligned}
    \end{equation}
    \item \textbf{MD-1}: the weights are fixed over the prediction window to the values computed using the Mahalanobis distance based on the one-step-behind optimal inputs, i.e., $\lambda^{\scriptscriptstyle[i]}(h) = \lambda^{\scriptscriptstyle[i]}_{\text{MD}}\big(u(k-1)\big), \: \text{for} \, i = 1,\dots,n$.
    \item \textbf{MD-2}: the weights are optimised within the MPC as a function of the inputs across the horizon, i.e., $\lambda^{\scriptscriptstyle[i]}(h) = \lambda^{\scriptscriptstyle[i]}_{\text{MD}}\big(u(h)\big), \: \text{for} \: i = 1,\dots,n$.
\end{itemize}
As a baseline controller against which to compare the proposed methods, a rule-based (\textbf{RB}) strategy is adopted: a constant supply temperature $T_0^\text{s} = 75$ that satisfies all constraints is used, and the storage flow is chosen as
\begin{equation}
    q_\text{\tiny TES}(h) = \begin{cases}
        -7.5 & c(h) < 0.125 \\
        7.5 & c(h) > 0.175 \\
        0 & \text{otherwise}
    \end{cases},
\end{equation}
such that charging and discharging of thermal energy is conducted when power production is cheap and expensive, respectively.

All simulations in the following are run on an 11th Gen Intel laptop with four i7 cores, 3.00GHz
clock speed, and 16GB of RAM. 
Optimisation problems are constructed using Casadi and solved using Ipopt.
Python source code is available at \url{https://github.com/SamuelMallick/dhs-ensemble}.

A three-day simulation, i.e., with $T=\frac{3 \cdot 86400}{\tau}$, is run with the load, electricity energy price, and constraint profiles shown in Figure \ref{fig:ext_inputs}.
\begin{figure}
    \centering
    \input{Images/tikz/external_inputs}
    \caption{ a) Thermal demand profiles. b) Electricity price trend. c) Temperature lower bounds: dashed for $T_0^\text{r}$, solid for $T_j^\text{s}$.}
    \label{fig:ext_inputs}
\end{figure}
The economic performance
\begin{equation}
    J = \sum_{k=0}^{T} \tau \cdot \sigma \cdot c(k) \cdot P_0(k),
\end{equation}
the constraint violations
\begin{equation}
    \begin{aligned}
        V = \sum_{k=0}^{T}\bigg( \sum_{j=1}^5 \Big (&\Big\|\max\big(0, \underline{T}_j^\text{s}(k)-T_j^\text{s}(k) \big)\Big\|^2 \\
         + & \Big\|\max\big(0,  T_j^\text{s}(k)-85\big)\Big\|^2 \Big) \\
        + & \Big\|\max\big(0,  \underline{T}_0^\text{r}(k)-T_0^\text{r}(k)\big)\Big\|^2 \\
        + & \Big\|\max\big(0, T_0^\text{r}(k)-75\big)\Big\|^2 \bigg),
    \end{aligned}
\end{equation}
and the average computation time for each approach (together with the standard deviation) are presented in Table \ref{tab:results}.
\begin{table}
\centering
\begin{tabular}{lccccc}
\toprule
 & RB & AV & LS & MD-1 & MD-2 \\
\midrule
$J$ [\euro] & 5965  & 5709 & 5547 & 5550 & \textbf{5495} \\
$V$ & \textbf{0} & 46.8 & 31.6 & \textbf{0} &\textbf{0} \\
$t$ [s] & - & \textbf{13.7$\pm$7.69} & 19.4$\pm$7.18 & 27.1$\pm$12.2 & 19.6$\pm$9.22 \\
\bottomrule
\end{tabular}
\caption{Performance of each control strategy: economic cost $J$, constraint violation $V$, and MPC average computation time $t$ with standard deviation.}
\label{tab:results}
\end{table}
\begin{figure*}
    \centering
    \input{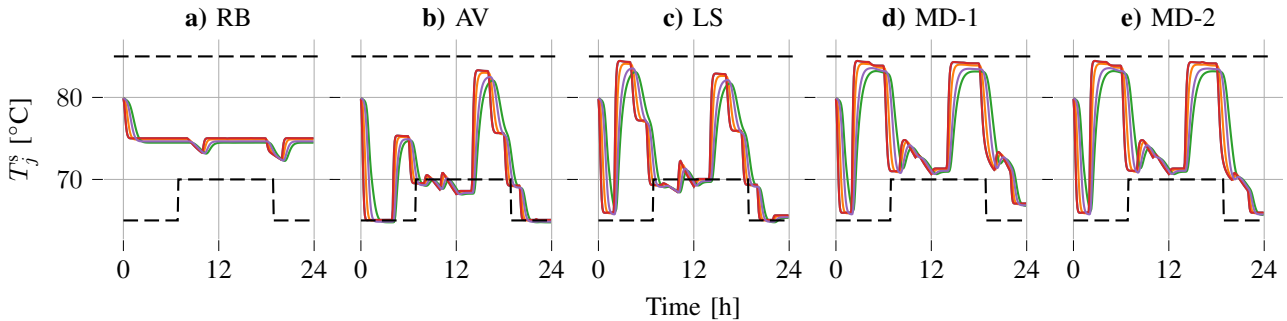}
    \caption{First day of simulation showing the five load supply temperatures $T_j^\text{s}$ along with the corresponding constraints (black dashed lines) under the different control strategies.}
    \label{fig:trajs}
\end{figure*}
The benefits of predictive control are clear from the improved economic cost for all MPC regulators with respect to the rule-based strategy. Among MPC regulators, the proposed approach, which optimises $\lambda^{\scriptscriptstyle[i]}$ over the prediction window as a function of inputs only (MD-2 \eqref{eq:lambdai}), achieves the best economic performance and the least violations of constraints. Furthermore, the computation burden required to optimise over the weights is comparable to that of the fixed-weighting approaches. Figure \ref{fig:trajs} shows the load supply temperatures $T_j^\text{s}$ for the first day of simulation. It can be observed that the AV and LS strategies exhibit constraint violations, whereas the Mahalanobis-based strategies do not. Moreover, the improvement in the prediction model for MD-2, where the ensemble weights are optimised over the prediction window, with respect to MD-1, where the weights are constant, is visible as the MD-2 controller reacts earlier to changes in future conditions, resulting in improved economic performance.

\begin{figure}
    \centering
    \input{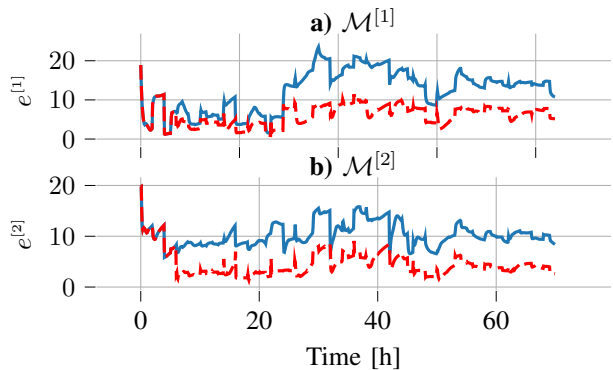}
    \caption{Output error $e^{\scriptscriptstyle[i]}$ for model in open-loop (solid blue) and with MHE-based state estimation (dashed red).}
    \label{fig:ouput_err}
\end{figure}

Finally, Figure \ref{fig:ouput_err} shows the one-step output error
\begin{equation}
    e^{\scriptscriptstyle[i]}(k) = \big\|y_{\text{p}}(k) - g^{\scriptscriptstyle[i]} \big(x^{\scriptscriptstyle[i]}(k), u(k)\big)\big\|_2
\end{equation}
for the data-based models during the same simulation, with $x^{\scriptscriptstyle[i]}(k)$ in \eqref{subeq:x0} estimated both from open-loop propagation of the model itself \eqref{eq:modeli} and from the proposed MHE scheme \eqref{eq:mhe}.
It can be observed that, thanks to the MHE-based state estimation, the output error is significantly reduced for both models $\mathcal{M}^{\scriptscriptstyle[1]}$ and $\mathcal{M}^{\scriptscriptstyle[2]}$, prior to being combined by the ensemble weighting.

\section{Conclusions}
This paper addresses optimal control of complex systems operating under multiple conditions using ensemble models. We propose a novel combination rule for ensemble models based on the statistical Mahalanobis distance, allowing the ensemble weights to evolve over the prediction window solely as functions of the inputs. In addition, a new observer for ensemble models based on MHE is developed to provide accurate MPC state initialisation. The proposed methodology is validated on a complex district heating system benchmark, demonstrating that the Mahalanobis distance-based combination rule outperforms both fixed and output-dependent weighting strategies in terms of economic performance and constraint satisfaction. The proposed MHE-based observer significantly enhances state estimation accuracy compared to an open-loop approach, leading to improved overall closed-loop performance. Future work will focus on integrating the proposed framework within an active learning setting to leverage MPC for informative data acquisition.
%\sam{QUESTO ARTICOLO HA CAMBIATO PER SEMPRE IL STATO DI ACADEMIA E IL MODO IN CUI LA GENTE LA GUARDA. IN PARTICOLARE SARÁ LA GOIELLO DI CORONA DELLE CARRIERA DI TUTTI QUELLI CHE SONO COINVOLTI!}

\section*{ACKNOWLEDGMENT}
The work of Laura Boca de Giuli, Alessio La Bella, and Riccardo Scattolini was carried out within the MICS Extended Partnership and received funding from Next-Generation EU (Italian PNRR - M4 C2, Invest 1.3 - D.D. 1551.11-10-2022, PE00000004). CUP MICS D43C22003120001.
The work of Samuel Mallick, Azita Dabiri, and Bart De Schutter was funded by the European Research Council (ERC) under the European Union's Horizon 2020 research and innovation programme (Grant agreement No. 101018826 - ERC Advanced Grant CLariNet).

\bibliographystyle{IEEEtran}
\bibliography{Bibliography}

\end{document}